\documentclass[%
 reprint,
 amsmath,amssymb,
 aps,
]{revtex4-1}

\usepackage{graphicx}
\usepackage{dcolumn}
\usepackage{bm}

\begin{document}

\preprint{APS/123-QED}

\title{Spacetime singularity resolution in Snyder noncommutative space}
\author{M. A. Gorji}
\homepage{m.gorji@stu.umz.ac.ir}
\author{K. Nozari}
\homepage{knozari@umz.ac.ir}
\affiliation{Department of Physics,
Faculty of Basic Sciences, University of Mazandaran,\\P. O. Box
47416-95447, Babolsar, Iran}

\author{B. Vakili}
\homepage{b-vakili@iauc.ac.ir} \affiliation{Department of Physics,
Chalous Branch,\\ Islamic Azad University, P.O. Box 46615-397,
Chalous, Iran}

\begin{abstract}
Inspired by quantum gravity proposal, we construct a deformed phase
space which supports the UV and IR cutoffs. We show that the
Liouville theorem is satisfied in the deformed phase space which
allows us to formulate the thermodynamics of the early Universe in
the semiclassical regime. Applying the proposed method to the Snyder
noncommutative space, we find a temperature dependent equation of
state which opens a new window for natural realization of inflation
as a phase transition from quantum gravity regime to the standard
radiation dominated era. Also we obtain finite energy and entropy
densities for the Universe, when at least the Weak Energy Condition
is satisfied. We show that there is a minimum size for the Universe
which is proportional to the Planck length and consequently
the Big Bang singularity is removed.\\
\begin{description}
\item[PACS numbers]
04.60.Bc, 98.80.Cq
\item[Key Words]
Quantum Gravity Phenomenology, Phase Transition, Big Bang Singularity
\end{description}
\end{abstract}
\maketitle
\section{Introduction}
The dynamics of the Universe in the Standard Big Bang scenario is
governed by the Einstein's general theory of relativity. If general
relativity is used to describe the observed Universe, the model
requires huge fine-tuned initial conditions \cite{IniCon}. Such an
initial conditions can be accommodated by the standard model, but of
course, they can not be explained in this framework. Inflationary
scenario, an accelerating phase before the nucleosynthesis era, can
resolve this problem in a novel way \cite{guth}. Such a scenario can
be realized, for instance, from a GUTs symmetry breaking phase
transition $SU(5)\rightarrow\,SU(3) \times{SU(2)}\times{U(1)}$
around $10^{15}\,\mbox{GeV}$ \cite{linde}. While inflation solves
the initial value problem, it is not a natural prediction of general
relativity. More precisely, general relativity is a classical theory
and is applicable at sub-Planckian curvatures. On the other hand,
inflation occurs in the quantum gravity regime and so it is
plausible to expect that the problem of initial condition will be
naturally addressed in the framework of full quantum gravity theory.
In the absence of a full quantum theory of gravity, we don't know
the exact dynamical equations governing on the early Universe.
Nevertheless, there are some candidates such as string theory and
loop quantum gravity which revealed some unknown aspects of the
quantum gravity. For instance, existence of a minimal length and a
minimal momentum which induce ultraviolet (UV) and infrared (IR)
cutoffs respectively, are common addresses of alternative candidates
for quantum gravity proposal \cite{QGML,GUPUV,DSR,NCG,NCGCS,Nicolini}.

Einstein equations for the Universe, including the Friedmann and
Raychaudhuri equations, can not be solved without supplementing an
equation of state. The equation of state is determined by
statistical analysis of the particles in the Universe. The question
then arises: is the equation of state in quantum gravity regime the
same as the low energy regime ones? We focus on this question in the
present study by considering a deformed phase space that includes
some phenomenological aspects of the quantum gravity proposal in
semiclassical regime. On the other hand, in the standard model of
cosmology the adiabatic expansion condition implicitly leads to the
Big Bang singularity when one reverses the expansion history. We
show here that quantum gravity effects modify the adiabatic
expansion condition leading trivially to a nonsingular early
Universe. Interestingly this resolves the large entropy density
problem too.

The structure of the paper is as follows: In Section 2, we introduce
a kinematical phase space $\tilde{\Gamma}$ which is consistent with
quantum gravity proposal and supports the existence of the UV and IR
cutoffs. In the dynamical level, we show that the Liouville theorem
is satisfied in the deformed phase space, which ensures that the
number of quantum states is invariant under the time evolution of
the system. In Section 3, we formulate the thermostatistics in
noncommutative phase space and we study some consequences of the
model in the early Universe cosmology. In Section 4, we apply the
proposed model to the Snyder noncommutative space. Section 5 is
devoted to the conclusions.

\section{The Deformed Phase Space}

The spacetime manifold structure is significantly affected by
quantum gravity effects in the high energy regime. All alternative
candidates of quantum gravity suggest some deformations of the
algebraic structure in such a way that the system under
consideration to be UV and IR regularized. These deformations can be
addressed properly through the modified Heisenberg commutation
relations. The Snyder noncommutative spaces are the well-known
example which can be realized from the modified commutation
relations and also are naturally UV/IR-regularized
\cite{NCG,Mignemi}. The generalized uncertainty principle is another
proposal which is suggested in the context of the string theory and
supports the existence of minimal length (UV cutoff) and minimal
momentum (IR cutoff) \cite{QGML,GUPUV}. Also, polymer quantization
is an effective approach to the loop quantum gravity \cite{LOOP}
which suggests the direct deformation to the phase space variables
through a process known as {\it polymerization}
\cite{corichi,corichi2,PLMRUR}. In what follows we formulate the
kinematics and dynamics of representative points in a phase space
with natural cutoffs.

\subsection{Kinematics}

Heisenberg algebra will be deformed in quantum gravity regime. The
deformed Heisenberg algebra leads to the deformed Poisson algebra
in the classical limit through a standard relation
$\frac{1}{i}[\hat{A}\,,\hat{B}]\rightarrow\{A\,,B\}$ \cite{units}.
The most general form of the noncanonical symplectic structure on
phase space $\tilde{\Gamma}$ of dimension $2D$ is
\begin{eqnarray}\label{commutation}
\{q_i\,,q_j\}=f_{ij}(q,p),\nonumber\\{\{q_i\,,p_j\}}=
g_{ij}(q,p),\\{\{p_i\,,p_j\}}=h_{ij}(q,p),\nonumber
\end{eqnarray}
where $i,j=1,2,...,D$. Here, $q$ and $p$ are the positions and
conjugate momenta respectively and $f_{ij}$, $g_{ij}$ and $h_{ij}$
are the differentiable functions which determine the deformed
Poisson algebra on $\tilde{\Gamma}$. Specifying particular forms
for these functions, one recovers the well-known noncommutative
algebras such as the Snyder and the Moyal algebras (see appendix A).

The above deformed Poisson brackets should have the same properties
as the usual Poisson brackets, that is, they should be
antisymmetric, bilinear, and satisfy the Leibnitz rule and the
Jacobi identity. Clearly, $f_{ij}$ and $h_{ij}$ should be totaly
antisymmetric $f_{ij}=-f_{ji}$ and $h_{ij}=-h_{ji}$ through the
antisymmetric property of the Poisson brackets. For two arbitrary
functions $U(\zeta_a)$ and $V(\zeta_a)$ where $\zeta_a= (q_i,p_i)$
with $a=1,2,...,2D$, the Poisson bracket reads
\begin{equation}\label{K-PB1}
\{U,V\}_{\zeta}=\{\zeta_a,\zeta_b\}\frac{
\partial{U}}{\partial{\zeta_a}}\frac{
\partial{V}}{\partial{\zeta_b}}\,.
\end{equation}
Expanding the above relation in terms of the phase space variables
$q$ and $p$ and using relations (\ref{commutation}) one obtains
\begin{eqnarray}\label{K_PB2}
\{U,V\}_{[q,p]}=g_{ij}\Big(\frac{\partial{U}}{
\partial{q_i}}\frac{\partial{V}}{\partial{p_j}}
-\frac{\partial{U}}{\partial{p_j}}\frac{
\partial{V}}{\partial{q_i}}\Big)\hspace{0.4cm}
\nonumber\\+f_{ij}\frac{\partial{U}}{\partial{
q_i}}\frac{\partial{V}}{\partial{q_j}}+h_{ij}
\frac{\partial{U}}{\partial{p_i}}\frac{
\partial{V}}{\partial{p_j}}.
\end{eqnarray}
Also, the Jacobi identity
\begin{eqnarray}\label{Jacobi-I}
\{U,\{V,W\}\}+\{V,\{W,U\}\}+\{W,\{U,V\}\}=0
\end{eqnarray}
is satisfied for any functions $U(q,p)$, $V(q,p)$, and $W(q,p)$ with
continuous second derivative. Substituting phase space variables $q$
and $p$ into the relation (\ref{Jacobi-I}) gives four independent
equations
\begin{eqnarray}\label{Jacobi-I1}
\{q_i,\{q_j,q_k\}\}+\{q_j,\{q_k,q_i\}\}+\{q_k,\{
q_i,q_j\}\}=0,\hspace{0.3cm}\nonumber\\{\{q_i,\{q_j,p_k\}\}}+\{
q_j,\{p_k,q_i\}\}+\{p_k,\{q_i,q_j\}\}=0,\hspace{0.3cm}\nonumber\\
{\{q_i,\{p_j,p_k\}\}}+\{p_j,\{p_k,q_i\}\}+\{p_k,\{
q_i,p_j\}\}=0,\hspace{0.2cm}\nonumber\\{\{p_i,\{p_j,p_k\}\}}+\{
p_j,\{p_k,p_i\}\}+\{p_k,\{p_i,p_j\}\}=0.\hspace{0.2cm}
\end{eqnarray}
Substituting from (\ref{commutation}) and using (\ref{K_PB2}), the
relations (\ref{Jacobi-I1}) give the constraints on the functions
$f_{ij}$, $g_{ij}$, and $h_{ij}$
\begin{eqnarray}\label{Jacobi-Func}
f_{im}\frac{\partial{f_{jk}}}{\partial{q_m}}+
g_{im}\frac{\partial{f_{jk}}}{\partial{p_m}}+
f_{jm}\frac{\partial{f_{ki}}}{\partial{q_m}}+
g_{jm}\frac{\partial{f_{ki}}}{\partial{p_m}}
\nonumber\\+f_{km}\frac{\partial{f_{ij}}}{
\partial{q_m}}+g_{km}\frac{\partial{f_{ij}}}{
\partial{p_m}}=0,\nonumber\\f_{im}\frac{
\partial{g_{jk}}}{\partial{q_m}}+g_{im}\frac{
\partial{g_{jk}}}{\partial{p_m}}-f_{jm}\frac{
\partial{g_{ik}}}{\partial{q_m}}-g_{jm}\frac{
\partial{g_{ik}}}{\partial{p_m}}\nonumber\\-
g_{mk}\frac{\partial{f_{ij}}}{\partial{q_m}}+
h_{km}\frac{\partial{f_{ij}}}{\partial{p_m}}=
0,\nonumber\\f_{im}\frac{\partial{h_{jk}}}{
\partial{q_m}}+g_{im}\frac{\partial{h_{jk}}}{
\partial{p_m}}+g_{mj}\frac{\partial{g_{ik}}}{
\partial{q_m}}-h_{jm}\frac{\partial{g_{ik}}}{
\partial{p_m}}\nonumber\\-g_{mk}\frac{
\partial{g_{ij}}}{\partial{q_m}}+h_{km}\frac{
\partial{g_{ij}}}{\partial{p_m}}=0,\nonumber
\\-g_{mi}\frac{\partial{h_{jk}}}{\partial{
q_m}}+h_{im}\frac{\partial{h_{jk}}}{\partial{
p_m}}-g_{mj}\frac{\partial{h_{ki}}}{\partial{
q_m}}+h_{jm}\frac{\partial{h_{ki}}}{\partial{
p_m}}\nonumber\\-g_{mk}\frac{\partial{
h_{ij}}}{\partial{q_m}}+h_{km}\frac{\partial{
h_{ij}}}{\partial{p_m}}=0.
\end{eqnarray}

We will see that the measure of the phase space $\tilde{\Gamma}$
is different from the measure of the usual phase space $\Gamma$.
To show this fact, consider a general noncanonical transformation
of the phase space
\begin{eqnarray}\label{transformation}
(Q,P)\rightarrow(q\,,\,p)\,,
\end{eqnarray}
where the variables $Q$ and $P$ obey the nondeformed Poisson algebra
on the phase space $\Gamma$
\begin{eqnarray}\label{Poisson}
\{Q_i,Q_j\}=0,\quad{\{Q_i,P_j\}}=\delta_{ij},\quad
\{P_i,P_j\}=0,\hspace{0.1cm}
\end{eqnarray}
while the variables $q=q(Q,P)$ and $p=p(Q,P)$ belong to the deformed
phase space $\tilde{\Gamma}$ and satisfy the deformed Poisson
algebra (\ref{commutation}). The Jacobian of transformation
(\ref{transformation}) in $2D$-dimensional classical phase space can
be expanded in terms of the Poisson brackets as \cite{Fityo,WF},
\begin{eqnarray}\label{Jacobian}
J(q,p)=\,\frac{\partial(q\,,\,p)}{\partial(Q,P)}=\hspace{5cm}
\\{\frac{1}{2^{D}D!}}\sum_{i_{1}...i_{2D}=1}^{2D}\epsilon_{
i_{1}..i_{2D}}\{X_{i_1},X_{i_2}\}...\{X_{i_{2D-1}},X_{
i_{2D}}\},\nonumber
\end{eqnarray}
where $\epsilon$ denotes the Levi-Civita symbol and $X_i$ denotes
the phase space variables so that for odd $i$ it is a coordinate
$q_i$ and for even $i$ it is a conjugate momenta $p_i$. The Jacobian
$J(q,p)$ induces the UV and IR cutoffs in the high and low energy
regimes respectively. The deformation to the measure of the phase
space $\tilde{\Gamma}$ can be obtained by means of the Jacobian
(\ref{Jacobian}) as
\begin{eqnarray}\label{QGWF}
\int_{\Gamma}(...)\,d\omega(Q,P)\,\longrightarrow\int_{
\tilde{\Gamma}}(...)\,\frac{d{\tilde{\omega}(q,p)}}{J(q,p)}
\end{eqnarray}
where $d\omega(Q,P)$ is the infinitesimal volume of the
$2D$-dimensional usual phase space $\Gamma$ and $d\tilde{
\omega}(q,p)/J(q,p)$ is its counterpart in the deformed phase space
$\tilde{\Gamma}$. It is important to note that the deformed phase
space volume $d\tilde{\omega}(q,p)/J(q,p)$ should be invariant under
the time evolution of the system. We consider this issue in the next
section and we show that the Liouville theorem is satisfied in the
deformed phase space $\tilde{\Gamma}$. In general, the phase spaces
$\Gamma$ and $\tilde{\Gamma}$ topologically may represent different
symplectic manifolds, but the problem arises is that how these two
manifolds coincide in the limit of the low energy regime? This
problem arises, for example, in polymer framework which can be
resolved by detailed analysis of the continuous limit of the
corresponding theory \cite{corichi,corichi2}. In the present study
these manifolds are topologically the same, though at the boundaries
they may behave differently because of, for instance, existence of
minimal length, minimal momentum and maximal momentum which may
affect the range of integrals in relation (\ref{QGWF}).

Although the ultimate form of the Jacobian $J(q,p)$ will be
specified just after formulating the full quantum gravity theory,
but effective theories to quantum gravity proposal have proposed
some candidates for this quantity
\cite{DOSGUP1,DOSGUP2L,DOSGUP3,DOSGUP4L,DOSDSR,PHG-DSR,DOSPLMR}.
Moreover, the deformation such as the relation (\ref{QGWF}), can
be deduced without demanding modified commutation relations. For
instance, the coherent states approach to the spacetime
noncommutativity provides a direct deformation to the phase space
which is equivalent to $J(q,p)^{-1}=e^{-\sigma{q^2}-\theta{p^2}}$
\cite{NCGCS,Nicolini} (see also Appendix A), where $\sigma$ and
$\theta$ are the noncommutative deformation parameters that
induce the IR and UV cutoffs respectively.

\subsection{Dynamics and the Liouville Theorem}
The next issue now is to consider the dynamics of the model. The
deformed measure
\begin{eqnarray}\label{NPSV}
\frac{d{\tilde{\omega}}(q,p)}{J(q,p)}\,,
\end{eqnarray}
determines the number of quantum states in the phase space
$\tilde{\Gamma}$. So, the deformed volume
$d\tilde{\omega}(q,p)/J(q,p)$ should be invariant under the time
evolution of the system to ensure that the Liouville theorem is
satisfied and consequently the number of microstates remains
unchanged.

Time evolution of any function of the phase space $U(q,p)$ in
Hamiltonian formalism can be represented by the Poisson brackets
\begin{eqnarray}\label{dynamics}
\frac{dU}{dt}=\{U,{\mathcal{H}}\},
\end{eqnarray}
where ${\mathcal{H}}(q,p)$ is the Hamiltonian of the system. The
equations of motion can be obtained from the relations
(\ref{K_PB2}) and (\ref{dynamics})
\begin{eqnarray}\label{EoM}
\dot{q}_i\;=\{q_i,{\mathcal{H}}\}\;=f_{ij}\,\frac{
\partial{{\mathcal{H}}}}{\partial q_j}+g_{ij}\,\frac{
\partial{{\mathcal{H}}}}{\partial p_j}\;,\nonumber\\
{\dot{p}_i}=\{p_i,{\mathcal{H}}\}\;=-g_{ji}\,\frac{
\partial{{\mathcal{H}}}}{\partial q_j}+h_{ij}\,\frac{
\partial{{\mathcal{H}}}}{\partial p_j}\;,
\end{eqnarray}
Consider an infinitesimal transformation of the phase space
variables $q_i$ and $p_i$
\begin{eqnarray}\label{Time-Trans}
q_i' & = & q_i + \delta q_i \;, \cr
p_i' & = & p_i + \delta p_i \;,
\end{eqnarray}
where $\delta q_i$ and $\delta p_i$ evolve through relations
(\ref{EoM}) as
\begin{eqnarray}\label{infin-evol}
\delta{q_i}=\Big(f_{ij}\frac{\partial{{\mathcal{H}}}}{\partial{q_j}}
+g_{ij}\frac{\partial{{\mathcal{H}}}}{\partial{p_j}}\Big)\delta{t}\;,
\nonumber\\{\delta{p_i}}=-\Big(g_{ji}\frac{\partial{{\mathcal{H}}}}{
\partial{q_j}}+h_{ji}\frac{\partial{{\mathcal{H}}}}{\partial{p_j}}
\Big)\delta{t}\;,
\end{eqnarray}
where we have used the fact that $h_{ij}=-h_{ji}$. An infinitesimal
deformed phase space volume evolves with time through the relations
(\ref{Time-Trans}) as
\begin{equation}\label{PSVTE}
d{\tilde{\omega}}(q',p')=\bigg|\frac{\partial(q_i',p_i')}
{\partial(q_i,p_i)}\bigg|\,d{\tilde{\omega}}(q,p)\;.
\end{equation}
From relations (\ref{Time-Trans}) we have
\begin{eqnarray}
\frac{\partial{q_i'}}{\partial{q_j}}=\delta_{ij}+\frac{
\partial\delta{q_i}}{\partial{q_j}},\qquad\frac{
\partial{q_i'}}{\partial{p_j}}=\frac{\partial\delta{q_i}}
{\partial{p_j}},\nonumber\\{\frac{\partial{p_i'}}
{\partial{q_j}}}=\frac{\partial\delta{p_i}}{\partial{q_j}},
\qquad\frac{\partial{p_i'}}{\partial{p_j}}=\delta_{ij}+
\frac{\partial\delta{p_i}}{\partial{p_j}}\,.
\end{eqnarray}
Using the above relations and up to the first order of $\delta{t}$
we have \cite{DOSGUP2L},
\begin{eqnarray}
\bigg|\frac{\partial(q_i',p_i')}{\partial(q_i,p_i)}\bigg|=
1+\Big(\frac{\partial{\delta{q_i}}}{\partial q_i}+\frac{
\partial{\delta{p_i}}}{\partial p_i}\Big)\,.
\end{eqnarray}
Substituting this relation in the relation (\ref{PSVTE}) one obtains
\begin{eqnarray}\label{DoPS1}
d{\tilde{\omega}}(q',p')=\bigg[1+\Big(\frac{\partial{f_{ij}}}{
\partial{q_i}}-\frac{\partial{g_{ji}}}{\partial{p_i}}\Big)
\frac{\partial{{\mathcal{H}}}}{\partial q_j}\delta{t}\hspace{2.5cm}
\nonumber\\+\Big(\frac{\partial{g_{ij}}}{\partial{q_i}}-\frac{
\partial{h_{ji}}}{\partial{p_i}}\Big)\frac{\partial{{\mathcal{H}}}}{
\partial p_j}\delta{t}\bigg]d{\tilde{\omega}}(q,p).\hspace{0.7cm}
\end{eqnarray}

In the next step we should consider the time evolution of the Jacobian
(\ref{Jacobian}). For small deviation from the usual Poisson algebra,
we have
\begin{eqnarray}\label{limit}
f_{ij}\ll 1,\qquad(g_{ij}-\delta_{ij})\ll 1,\qquad\,h_{ij}\ll 1,
\end{eqnarray}
independent of the explicit form of these functions. The above
conditions ensure that the noncanonical sympelctic structure
(\ref{commutation}) reduces to the usual canonical ones in the low
energy limit. In this limit, the Jacobian (\ref{Jacobian}) can be
approximated as \cite{Fityo}
\begin{eqnarray}\label{Jacobianapprox}
J(q,p)=\prod_{i=1}^D\{q_i,p_i\}=\prod_{i=1}^D\,g_{ii}\approx{
1+\sum_{i=1}^D\,\big(g_{ii}-1\big)}.\hspace{0.4cm}
\end{eqnarray}
The time evolution of the above Jacobian can be obtained through
relations (\ref{Time-Trans}) as
\begin{eqnarray}\label{Time-Jacobianapprox}
J(q',p')=\prod_{i=1}^D\{q_i',p_i'\}=\prod_{i=1}^D\Big(\{
q_i,p_i\}+\{q_i,\delta{p_i}\}\hspace{1cm}\nonumber\\+\{\delta{
q_i},p_i\}+\{\delta{q_i},\delta{p_i}\}\Big).\hspace{0.7cm}
\end{eqnarray}
The last term $\{\delta{q_i},\delta{p_i}\}$ is the second order of
$\delta{t}$ and can be ignored. Then, up to the first order of
$\delta{t}$ we find
\begin{eqnarray}\label{Time-Jacobianapprox2}
J(q',p')=\prod_{i=1}^D\{q_i,p_i\}+\prod_{i=1}^D\Big(\{
q_i,\delta{p_i}\}+\{\delta{q_i},p_i\}\Big).\hspace{0.4cm}
\end{eqnarray}
The first term in the right hand side of the above relation
coincides with relation (\ref{Jacobianapprox}). Substituting from
relations (\ref{infin-evol}), to first order of $\delta{t}$ the
relation (\ref{Time-Jacobianapprox2}) becomes
\begin{eqnarray}\label{Time-Jacobianapprox3}
J(q',p')=J(q,p)+\prod_{i=1}^D\bigg[\,g_{ik}\Big(\frac{
\partial{f_{kj}}}{\partial{q_i}}\,-\,\frac{\partial{g_{ji}}}{
\partial{p_k}}\Big)\frac{\partial{{\mathcal{H}}}}{
\partial{q_j}}\hspace{2cm}\nonumber\\+g_{ik}\Big(\frac{
\partial{g_{kj}}}{\partial{q_i}}-\frac{\partial{h_{ji}}}{
\partial{p_k}}\Big)\frac{\partial{{\mathcal{H}}}}{
\partial{p_j}}\hspace{1.6cm}\nonumber\\-f_{ik}\Big(\frac{
\partial{g_{ij}}}{\partial{q_k}}\frac{\partial{
{\mathcal{H}}}}{\partial{q_j}}+\frac{\partial{h_{ji}}}{
\partial{q_k}}\frac{\partial{{\mathcal{H}}}}{\partial{p_j}}
\Big)\hspace{1.1cm}\nonumber\\+h_{ik}\Big(\frac{\partial{f_{kj}}}{
\partial{p_i}}\frac{\partial{{\mathcal{H}}}}{\partial{
q_j}}+\frac{\partial{g_{jk}}}{\partial{p_i}}\frac{\partial{
{\mathcal{H}}}}{\partial{p_j}}\Big)\bigg]\delta{t},
\hspace{0.5cm}\nonumber
\end{eqnarray}
where we have used relation (\ref{Jacobianapprox}). In the light of
the relation (\ref{limit}), one can neglect the second order terms
\begin{eqnarray}\label{limitSO}
(g_{ij}-\delta_{ij})\times\,f_{ij},\qquad(g_{ij}-\delta_{ij})
\times\,h_{ij},\qquad f_{ij}\times\,h_{ij}.\hspace{0.7cm}
\end{eqnarray}
So, the last two terms can be neglected and the above Jacobian
reduces to the following relation
\begin{eqnarray}\label{Time-Jacobianapprox4}
J(q',p')\approx J(q,p)+\Big(\frac{\partial{f_{ij}}}{\partial{
q_i}}-\frac{\partial{g_{ji}}}{\partial{p_i}}\Big)\frac{
\partial{{\mathcal{H}}}}{\partial q_j}\delta{t}\hspace{1cm}
\nonumber\\+\Big(\frac{\partial{g_{ij}}}{\partial{q_i}}-
\frac{\partial{h_{ji}}}{\partial{p_i}}\Big)\frac{\partial{
{\mathcal{H}}}}{\partial p_j}\delta{t},\hspace{0.5cm}
\end{eqnarray}
which after some manipulation becomes
\begin{eqnarray}\label{Time-Jacobianapprox5}
\frac{J(q',p')}{J(q,p)}=1+J^{-1}(q,p)\bigg[\Big(\frac{
\partial{f_{ij}}}{\partial{q_i}}-\frac{\partial{g_{ji}}}{
\partial{p_i}}\Big)\frac{\partial{{\mathcal{H}}}}{
\partial q_j}\hspace{1cm}\nonumber\\+\Big(\frac{\partial{
g_{ij}}}{\partial{q_i}}-\frac{\partial{h_{ji}}}{\partial{
p_i}}\Big)\frac{\partial{{\mathcal{H}}}}{\partial p_j}
\bigg]\delta{t}.\hspace{0.5cm}
\end{eqnarray}
The inverse of the Jacobian can be approximated through the relation
(\ref{Jacobianapprox}) as
\begin{eqnarray}\label{Jacobianinverse}
J^{-1}(q,p)\approx\,1-\sum_{i=1}^D\,\big(g_{ii}-1\big),
\hspace{0.7cm}
\end{eqnarray}
where we have used the relation (\ref{limit}). Substituting inverse
of the Jacobian (\ref{Jacobianinverse}) and again neglecting the
second order terms, one gets
\begin{eqnarray}\label{Time-Jacobianapprox6}
\frac{J(q',p')}{J(q,p)}=1+\Big(\frac{\partial{f_{ij}}}{
\partial{q_i}}-\frac{\partial{g_{ji}}}{\partial{p_i}}\Big)
\frac{\partial{{\mathcal{H}}}}{\partial q_j}\delta{t}
\hspace{1cm}\nonumber\\+\Big(\frac{\partial{g_{ij}}}{
\partial{q_i}}-\frac{\partial{h_{ji}}}{\partial{p_i}}\Big)
\frac{\partial{{\mathcal{H}}}}{\partial p_j}\delta{t}.
\hspace{0.5cm}
\end{eqnarray}
From relations (\ref{Time-Jacobianapprox6}) and (\ref{DoPS1})
we have
\begin{eqnarray}\label{invariant}
\frac{d{\tilde{\omega}}(q',p')}{J(q',p')}=\frac{d{\tilde{
\omega}}(q,p)}{J(q,p)}\,,
\end{eqnarray}
which ensures that the deformed phase space volume (\ref{NPSV}) is
invariant under time evolution of the system and consequently the
Liouville theorem is satisfied in the deformed phase space
$\tilde{\Gamma}$. This result is very essential for our forthcoming
arguments. We note that our general results obtained in this section
include the results obtained in special cases studied previously
\cite{DOSGUP2L,DOSGUP4L}.

\section{Thermostatistics}
The volume of the phase space determines the number of microstates
in the semiclassical regime and according to the Liuoville theorem
it should be invariant under the time evolution. Now we are able to
formulate the statistical mechanics in noncommutative phase space
since the deformed density of states (\ref{QGWF}) is invariant under
the time evolution through the relation (\ref{invariant}). Moreover,
one should also be careful about the definition of the bosons and
fermions due to the loss of the Lorentz invariance in a
noncommutative spacetime \cite{LINC}. Here we suppose that fermions
and bosons are defined in the same way as in the standard quantum
mechanics but within the \emph{coherent state picture} of
noncommutativity which considers a particle as a smeared objects
rather than to being a point-like particle. In other words, bosons
and fermions save their quantum mechanical properties as in the
standard quantum mechanics but the effect of noncommutativity of
space is implemented by a substitution rule: the point-like
structure of these particles is assumed to be replaced by a smeared,
Gaussian profile. In this formalism, the particle mass $M$, instead
of being completely localized at a point, is distributed throughout
a region of linear size $\sqrt{\theta}$ (see part 2 of Appendix A).
The implementation of this argument leads to the substitution of a
position Dirac-delta function (which describes point-like
structures) with a Gaussian profile describing smeared structures
\cite{NCGCS,Nicolini,Nozmeh}. As has been shown in Ref. \cite{Wung},
the space noncommutativity enhances the negative statistical
correlation between fermions and enhances the positive statistical
correlation between bosons. Also, there are residual "attraction
potential" between bosons and residual "repulsion potential" between
fermions in the high temperature limit. So, in a noncommutative
space the usual knowledge in statistical mechanics is still true,
say the Bose-Einstein or Fermi-Dirac distributions with a modified
density of states for smeared particles. With these points in mind,
in which follows we treat the thermostatistics of bosons and
fermions in this setup.

\subsection{The Method}
The issue of the noncommutativity can be included in the phase space
by two equivalent pictures \cite{NCAPP}: {\it i}) Working with the
deformed commutation relation, such as (\ref{commutation}), together
with the non-deformed Hamiltonian function, {\it ii}) Finding
canonical variables on the noncommutative phase space which satisfy
the commutative algebra but the Hamiltonian function now gets
modified to ensure that the Hamilton's equations (\ref{EoM}) being
the same in the two pictures. Mathematically, these two pictures are
related to each other by the Darboux transformation. According to
the Darboux theorem, it is always possible to find canonical
coordinates on the symplectic manifold which satisfy commutative
algebra. So, it is always possible to find a transformation that
transforms any noncommutative Poisson algebra such as
(\ref{commutation}) to the commutative ones \cite{Arnold}. Of
course, the Hamiltonian function gets modified when one transforms
the noncommutative algebra to the commutative ones to ensure that
the trajectories on the phase space remain the same in the two
pictures. However, working within the first picture is more
significant in statistical mechanics since in this picture
noncommutativity only affects the number of microstates through the
deformed density of states.

The number of particles $\mathrm{N}$ and pressure $\mathrm{P}$ of
a statistical system with volume $V$ at temperature $T$ is given by
the standard definitions
\begin{equation}\label{Ndef}
{\mathrm{N}}=\sum_{\varepsilon}\big(z^{-1}\,e^{\varepsilon/T}
\,\mp\,1\big)^{-1}\,,
\end{equation}
and
\begin{equation}\label{Pdef}
{\mathrm{P}}V=\mp\,T\sum_{\varepsilon}\ln\big(1\mp\,z\,
e^{-\varepsilon/T}\big)\,,
\end{equation}
respectively, where $z$ is the fugacity of the system and signs
$(-)$ and $(+)$ hold for bosons and fermions respectively. The
energy of the microstates $\varepsilon$ should be determined only by
quantized theory. In usual statistical mechanics, $\varepsilon$ is
the solution of the Schr\"{o}dinger equation. Here it should be a
solution of the full quantum gravity equations for the corresponding
statistical system. But, one can replace summation over
$\varepsilon$ by the integral over all phase space variables by
means of the density of states (\ref{QGWF}) as
$\sum_{\varepsilon}\rightarrow\,\frac{1}{(2\pi)^3}
\int_{V}\int\frac{d^3{q}\,d^3{p}}{J(q,p)}$, where $V$ is the volume
of the corresponding statistical system. In the early Universe, all
the particles effectively are relativistic and the Hamiltonian
simplifies to ${\mathcal{H}}(p)=p$ (where $p$ is the norm of the
vector $p_{i}$). We set also $z=1$ (the chemical potential to be
zero) as usually one assumes. Now, the number of particles and
pressure in quantum gravity regime can be obtained from the
relations (\ref{Ndef}) and (\ref{Pdef}) as
\begin{eqnarray}\label{N}
{\mathrm{N}}_{\mp}=\frac{g_{_{\mp}}}{(2\pi)^3}\,\int_V\int\,
\big(e^{p/T}\,\mp\,1\big)^{-1}\frac{d^3{q}\,d^3{p}}{J(q,p)},
\end{eqnarray}
\begin{eqnarray}\label{P}
{\mathrm{P}}_{\mp}=\mp\frac{g_{_{\mp}}}{(2\pi)^3}\,\frac{T}{V}
\int_V\int\,\ln\big(1\mp\,e^{-p/T}\big)\frac{d^3{q}\,d^3{p}}{J(q,p)}\,,
\end{eqnarray}
where $g_{_{-}}$ and $g_{_{+}}$ are the number of relativistic
degrees of freedom for bosons and fermions respectively. In the
above relations, the sign $(-)$ and $(+)$ hold for the bosons
and fermions respectively. The usual results can be recovered
by setting $J=1$ which is corresponding to the identity
transformation with $q=Q$ and $p=P$ in relation
(\ref{transformation}). The total number of particles and total
pressure is given by
\begin{eqnarray}\label{Ntot}
{\mathrm{N}}={\mathrm{N}}_{_{-}}+{\mathrm{N}}_{_{+}}\,,
\end{eqnarray}
and
\begin{eqnarray}\label{Ptot}
{\mathrm{P}}={\mathrm{P}}_{_{-}}+{\mathrm{P}}_{_{+}}\,.
\end{eqnarray}

The entropy density $s$ and the
energy density $\rho$ of the system can be obtained from the
definitions
\begin{eqnarray}\label{definition}
s(T)=\frac{\partial{\mathrm{P}}}{\partial{T}}\,,\hspace{1.2cm}
\rho(T)=T^2\frac{\partial}{\partial T}\bigg(\frac{\mathrm{P}}{T}
\bigg)\,.
\end{eqnarray}
Now, the semiclassical statistical consideration is completed and
one can obtain any thermodynamical quantities in noncommutative
phase spaces through the relations (\ref{Ntot}), (\ref{Ptot}),
and (\ref{definition}).

\subsection{Cosmological Implications}
Before considering particular examples of the noncommutative phase
space, we study some implications of our setup on the thermodynamics
of the early Universe.

The dynamics of the Universe in the standard cosmology is given by
the Einstein equations, the so-called Friedmann and Raychaudhuri
equations,
\begin{eqnarray}\label{freidmann}
\Big(\frac{\dot{a}}{a}\Big)^2+\frac{k}{a^2}=\frac{8\pi{G}}{3}
\rho\,,
\end{eqnarray}
\begin{eqnarray}\label{raychadhouri}
\frac{\ddot{a}}{a}=-\frac{4\pi{G}}{3}(\rho+3{\mathrm{P}})\,,
\end{eqnarray}
where $a(t)$ is the scale factor, $\rho$ and ${\mathrm{P}}$ are the
energy density and pressure respectively, and a dot denotes
derivative with respect to the cosmic time. Here $k$ marks the
spatial curvature which is normalized to zero, $1$ and $-1$ for
flat, closed and open Universes, respectively. Furthermore, an
equation of state should be supplemented to complete this set of
equations. In fact, an equation of state parameter of the form
$\mathrm{P}=\mathrm{P}(\rho,s)$ determines whether the Universe is
accelerating or decelerating, through the Raychaudhuri equation. In
principle, the equation of state should be obtained from the
statistical considerations of the particles in the early Universe.
So, the question is that whether equation of state remains unchanged
in the limit of high temperature? From the relation (\ref{P}) and
the definition (\ref{definition}) for the energy density, it is
clear that the equation of state changes when one includes quantum
gravity effects. We find such a modification to the equation of
state in Snyder spaces in the next section.

Furthermore, it is also important to note that the entropy density
(\ref{definition}) now changes since the pressure is modified
through the relation (\ref{P}). Consequently, the adiabatic
condition
\begin{eqnarray}\label{adiabatic}
S=s\,a^3=\mbox{cte.}\,,
\end{eqnarray}
where $S$ is the total entropy of the Universe, gets modified in
quantum gravity regime. Such a modification to the entropy density
removes the Big Bang singularity in a fascinating manner. We will
see this feature explicitly in the case of Snyder noncommutative
space in the next section.

\section{The Snyder Universe}
The Snyder noncommutative spacetime was firstly introduced by Snyder
\cite{NCG}. The corresponding noncommutative phase space has
recently been developed in Ref. \cite{Mignemi} by means of an
appropriate structure with the following noncommutative commutation
relations (as has been shown by Mignemi in \cite{Mignemi}, this is
actually the Snyder space on a sphere)
\begin{eqnarray}\label{snyderalgebra}
\{q_i,q_j\}=\beta^2\,J_{ij},\qquad\{p_i,p_j\}=\alpha^2\,J_{ij},
\nonumber\\{\{q_i,p_j\}}=\delta_{ij}+\alpha^2\,q_i
q_j+\beta^2\,p_i p_j+2\alpha\beta\,p_i q_j,
\end{eqnarray}
where $i,j=1,2,...,D$ and $J_{ij}=q_ip_j-q_jp_i$ are the generators
of the rotation in $D$ dimensions. The deformation parameters
$\alpha$ and $\beta$ induce the IR and UV cutoffs respectively. We
need the Jacobian (\ref{Jacobian}) corresponding to the Snyder
algebra (\ref{snyderalgebra}) to study the thermodynamics in this
framework by using the relations (\ref{Ntot}), (\ref{Ptot}), and
(\ref{definition}). In Appendix A we have calculated the Jacobian
for the Snyder space which is
\begin{eqnarray}\label{JacobianSnyder}
J(q,p)=1+\,3|\alpha{\bf q}+\beta{\bf p}|^2.
\end{eqnarray}
where we have set $D=3$ for a single-particle states and ${\bf q}$
and ${\bf p}$ are the 3-vectors associated to the $q_i$ and $p_i$
respectively. In Appendix A we have shown that the Jacobian
(\ref{JacobianSnyder}) also supports the other approaches to the
noncommutativity such as the coherent state approach
\cite{NCGCS,Nicolini}. The Jacobian (\ref{JacobianSnyder}) contains
the UV/IR mixing effect which is a common feature of the
noncommutative spaces. Both the IR and UV cutoffs are essential for
the renormalization of the quantum fields in curved spaces.
Substituting the Jacobian (\ref{JacobianSnyder}) in the relation
(\ref{P}) gives the pressure for the bosons and fermions as
\begin{widetext}
\begin{eqnarray}\label{SnyderP2}
{\mathrm{P}}_{\mp}=\mp\frac{g_{_{\mp}}}{(2\pi)^3}\,\frac{T}{V}\int_V\int\,
\ln\big(1\mp\,e^{-p/T}\big)\times\Big(1-3\alpha^2q^2-3\beta^2p^2-6\alpha
\beta{\bf q.p}\Big)\,d^3{q}\,d^3{p}\,\nonumber\\
=\mp\frac{g_{_{\mp}}}{(2\pi)^3}\,\frac{T}{V}\Bigg(\int_V\,d^3q\times\int\,
\ln\big(1\mp\,e^{-p/T}\big)d^3p-3\alpha^2\int_V\,q^2\,d^3q\times\int
\ln\big(1\mp\,e^{-p/T}\big)d^3p\nonumber\\-3\beta^2\int_V\,d^3q\times\int
\ln\big(1\mp\,e^{-p/T}\big)p^2\,d^3p-6\alpha\beta\int_V\int{\bf q.p}\,
\ln\big(1\mp\,e^{-p/T}\big)\,d^3{q}\,d^3{p}\Bigg),
\end{eqnarray}
\end{widetext}
where we have expanded the Jacobian up to the second order of the
deformation parameters $\alpha$ and $\beta$. Only the last term in
the right hand side of the above relation includes both of the
deformation parameters $\alpha$ and $\beta$. Now we calculate this
UV/IR mixing term. Writing the 3-vectors ${\bf q}$ and ${\bf p}$ in
the spherical coordinates as ${\bf q}=(q,\theta_1,\varphi_1)$ and
${\bf p}=(p,\theta_2,\varphi_2)$ with $q=|{\bf q}|$ and $p=|{\bf
p}|$, the inner product will be ${\bf
q.p}=qp\big(\cos\theta_1\,\cos\theta_2+
\sin\theta_1\sin\theta_2\cos(\varphi_1-\varphi_2)\big)$ and the last
term in the relation (\ref{SnyderP2}) becomes
\begin{widetext}
\begin{eqnarray}
-6\alpha\beta\int_V\int{\bf q.p}\,\ln\big(1\mp\,e^{-p/T}\big)\,
d^3{q}\,d^3{p}=-6\alpha\beta\int_0^{\infty}dp\,p^3\ln\big(1\mp\,
e^{-p/T}\big)\int_{-1}^{+1}d(\cos\theta_2)\int_{0}^{2\pi}d\varphi_2
\,\nonumber\\{\times}\,\int_{0}^{R}d{q}\,q^3\int_{-1}^{+1}
d(\cos\theta_1)\int_{0}^{2\pi}d\varphi_1\big(\cos\theta_1\,
\cos\theta_2+\sin\theta_1\sin\theta_2\cos(\varphi_1-\varphi_2)
\big)=0\,,
\end{eqnarray}
\end{widetext}
where we have used the integrals $\int_{-1}^{+1}d(\cos\theta_1)
\cos\theta_1=0$ and
$\int_{0}^{2\pi}d\varphi_1\cos(\varphi_1-\varphi_2) =0$. So, the
last term in the relation (\ref{SnyderP2}) vanishes. The integral
over coordinates simply gives $\int_{V}d^3q=V$,
$\int_{V}q^2\,d^3q=\frac{4\pi}{5}R^5=\frac{3}{5}\Big(\frac{3}{4\pi}\Big)^{2/3}V^{5/3}$
where $R$ is the radius corresponding to the volume of the system
under consideration. Performing the integrals over momentums by
using the identity $\int_{0}^{\infty}\frac{x^{n-1}\,dx}{e^x+1}=
\big(1-\frac{1}{2^{n-1}}\big)\,\int_{0}^{\infty}\frac{x^{n-1}\,dx
}{e^x-1}$ gives the pressure for the bosons and fermions
respectively as
\begin{eqnarray}\label{SnyderP3}
{\mathrm{P}}_{-}=g_{_{-}}\frac{\pi^2T^4}{90}\bigg(1-\frac{9}{5}
\Big(\frac{3}{4\pi}\Big)^{\frac{2}{3}}\alpha^2\,V^{\frac{2}{3}}-
\frac{24}{7}\beta^2\pi^2T^2\bigg),\hspace{0.5cm}\\{\mathrm{P}}_{+}
=\frac{7}{8}g_{_{+}}\frac{\pi^2T^4}{90}\bigg(1-\frac{9}{5}\Big(
\frac{3}{4\pi}\Big)^{\frac{2}{3}}\alpha^2\,V^{\frac{2}{3}}-\frac{
186}{49}\beta^2\pi^2T^2\bigg),\nonumber
\end{eqnarray}
where as usual $g_{_{-}}$ and $g_{_{+}}$ are the number of
relativistic degrees of freedom for bosons and fermions
respectively. The natural choice for the IR deformation parameter is
the square root of the cosmological constant with $\alpha\sim
10^{-24}{\mbox{cm}}^{-1}$, and for the UV deformation parameter
$\beta=\beta_0\,l_{_P}=\beta_0\,T_{_P}^{-1}$ is relevant to ensure
that the UV effects only become important around the Planck scale.
The numerical constant $\beta_0\sim\,{\mathcal{O}}(1)$ should be
fixed only with experiment \cite{experiment}. The total pressure of
the system can be obtained from the relations (\ref{Ptot}) and
(\ref{SnyderP3}) as
\begin{eqnarray}\label{PtotIRUV}
{\mathrm{P}}(T)=g_{\star}\frac{\pi^2{T^4}}{90}
\bigg(1-\frac{9}{5}\Big(\frac{3}{4\pi}\Big)^{\frac{2}{3}}
\alpha^2 V^{\frac{2}{3}}-\chi(T/T_{_P})^2\bigg),\hspace{0.5cm}
\end{eqnarray}
where $g_{\star}=\big(g_{_{-}}+\frac{7}{8}g_{_{+}}\big)$ and
$\chi=\frac{24{\pi^2}}{7}\Big(\frac{g_{_{-}}+\frac{31}{32}
g_{_{+}}}{g_{_{-}}+\frac{7}{8}g_{_{+}}}\Big)\,\beta_0^2$. We are
interested in the early Universe implications and the IR term in the
relation (\ref{PtotIRUV}) is negligible in the high temperature
limit. Nevertheless, the very small IR effects play an essential
role for the renormalization of the quantum fields in the curved
space. Therefore, the total pressure in the high temperature limit
will be
\begin{eqnarray}\label{PtotNC}
{\mathrm{P}}(T)=g_{\star}\frac{\pi^2{T^4}}{90}
\,\Big(1-\chi(T/T_{_P})^2\Big).
\end{eqnarray}
The corresponding energy density can be obtained through the
definition (\ref{definition}) as
\begin{eqnarray}\label{rhoNC}
\rho(T)=g_{\star}\frac{\pi^2{T^4}}{30}\,
\Big(1-\frac{5}{3}\chi(T/T_{_P})^2\Big).
\end{eqnarray}

\subsection{Energy Conditions and Equation of State Parameter}

The results (\ref{PtotNC}) and (\ref{rhoNC}) show that the pressure
and energy density are not always positive definite in Snyder space.
So one can not use these results in the Einstein equations
(\ref{freidmann}) and (\ref{raychadhouri}) without considering the
energy conditions for them. While both of the pressure and energy
density get negative values in the limit of high temperature
$T\rightarrow\,\infty$, the energy conditions give a correct picture
for these thermodynamical quantities. In table \ref{tab:table1}, we
represent the temperature intervals for the validity of the Dominant
Energy Condition (DEC), Weak Energy Condition (WEC), and Strong
Energy Condition (SEC).
\begin{table}
\begin{tabular}{|c|c|c|c|c|c|}
\hline Energy conditions & Temperature\\
\hline DEC, WEC, SEC & $0<T<\frac{1}{\sqrt{2\chi}}\,T_{_P}$ \\ \hline
WEC, SEC & $0<T<\sqrt{\frac{3}{5\chi}}\,T_{_P}$ \\ \hline
SEC & $0<T<\sqrt{\frac{2}{3\chi}}\,T_{_P}$ \\ \hline
$\rho<0,\quad{\mathrm{P}}\leq0$ &
$T\geq\frac{1}{\sqrt{\chi}}\,T_{_P}$\\ \hline
\end{tabular}
\caption{\label{tab:table1} Energy conditions in Snyder
noncommutative space}
\end{table}
The pressure always is positive for the temperature
$T<\frac{1}{\sqrt{\chi}}$, so all of the intervals in the table
\ref{tab:table1} are the subset of the domain of validity of DEC,
WEC, and SEC. In what follows, we consider the thermodynamics of the
Snyder space in different energy conditions separately.

\begin{figure}
\flushleft\leftskip+3em{\includegraphics[width=2.5in]{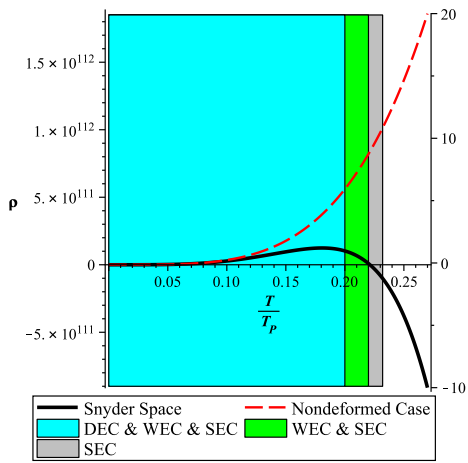}}
\hspace{3cm}\caption{\label{fig:1} The energy density versus the
temperature in the Snyder space. The deformation parameter is taken
to be $\beta_0=1$ and
the number of relativistic degrees of freedom for the bosons and
fermions are taken as $g_{_{-}}=2$ and $g_{_{+}}=10$ respectively.
In the limit of high temperature, the quantum gravity effects become
efficient and the energy density decreases and consequently the
infinite energy density in the Big Bang model disappears. The domain
of the validity of the energy conditions are shown with cyan, green
and silver colors.}
\end{figure}

The energy density as a function of the temperature is shown in
figure \ref{fig:1}. Interestingly, the infinite energy density which
appears in the standard Big Bang model now is removed and the energy
density behaves differently in the high energy regime. We will show
that this result emerges because the Big Bang singularity is removed
in the Snyder space.

\begin{figure}
\flushleft\leftskip+3em{\includegraphics[width=2.5in]{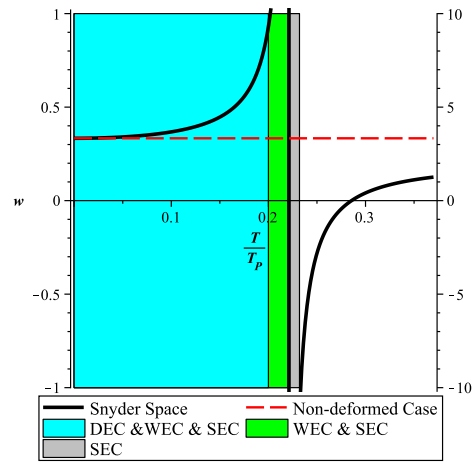}}
\hspace{3cm}\caption{\label{fig:2} The equation of state parameter
$w$ which becomes a function of temperature in the Snyder space. In
the cyan region, all of the DEC, WEC, and SEC energy conditions are
satisfied and the equation of state parameter is restricted to the
range $\frac{1}{3}\leq{w}\leq{1}$. In the green region, where WEC
and SEC are satisfied, $w\in[1,+\infty)$. In the silver region when
only the SEC is satisfied, $w\in(-\infty,-1]$. In this region the
condition $w=-1$ is possible, however this holds for negative values
of the energy density and therefore doesn't give an accelerating
expansion.}
\end{figure}

The equation of state parameter $w={\mathrm{P}}/\rho$ becomes
\begin{eqnarray}\label{omegaNC}
w(T)=\,\frac{1-\chi(T/T_{_P})^2}{3-5\chi(T/T_{_P})^2}\,.
\end{eqnarray}
The above relation correctly reduces to the usual radiation
dominated case in the limit of low temperatures $w(T\rightarrow 0)=
\,\frac{1}{3}$. This form of the equation of state is very similar
to the one obtained in Ref. \cite{NCInf}, where the authors proposed
a noncommutative inflation in the framework of the varying speed of
light theories. Clearly, the condition $w=-1$ is now possible for
the temperature $T=\sqrt{\frac{2}{3\chi}} \,T_{_P}$ where only the
SEC is satisfied. While the SEC is satisfied in this temperature,
the energy density (\ref{rhoNC}) becomes negative for this
temperature (see also figure \ref{fig:1}). In this situation the
condition $w=-1$ doesn't give accelerating expansion since gravity
is always attractive when the SEC is satisfied. The temperature
evolution of the equation of state parameter is shown in figure
\ref{fig:2}. The equation of state parameter varies in the range of
$\frac{1}{3}\leq{w}\leq{1}$ when all of the DEC, WEC, and SEC
simultaneously are satisfied. It varies in the range
$w\in[1,+\infty)$ when both the WEC and SEC are satisfied. The
negative values for the equation of state parameter
$w\in(-\infty,-1]$ are allowed when only the SEC is satisfied.

\subsection{Entropy Density and Big Bang Singularity}

\begin{figure}
\flushleft\leftskip+3em{\includegraphics[width=2.5in]{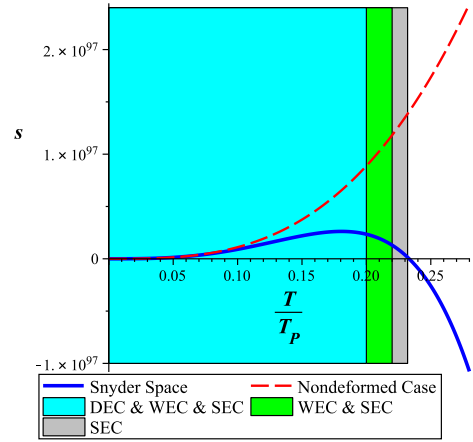}}
\hspace{3cm}\caption{\label{fig:3} Entropy density variation versus
the temperature in the Snyder space. The unusual
behavior of the entropy density simultaneously resolves the large
entropy density problem and the Big Bang singularity.}
\end{figure}

As we have seen previously, the entropy density modifies when one
considers the quantum gravity effects in the thermodynamics of the
early Universe through the relations (\ref{P}) and (\ref{definition}).
This different entropy density significantly changes the adiabatic
condition (\ref{adiabatic}) which determines the temperature
evolution of the Universe. The entropy density in high temperature
limit can be obtained by substituting the relation (\ref{PtotNC})
in (\ref{definition})
\begin{eqnarray}\label{entropyNC}
s(T)=g_{\star}\frac{2\pi^2{T^3}}{45}
\,\Big(1-\frac{3}{2}\chi(T/T_{_P})^2\Big).
\end{eqnarray}
Note that for $w=-1$ we find $s(\sqrt{\frac{2}{3\chi}}\,T_{_P})=0$
since definitions (\ref{definition}) indicate that
$s=\frac{\rho+{\mathrm{P}}}{T}$. The temperature behavior of the
entropy density is shown in figure \ref{fig:3}.

\begin{figure}
\flushleft\leftskip+3em{\includegraphics[width=2.5in]{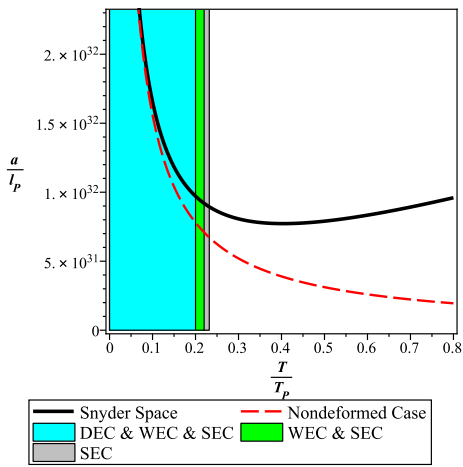}}
\hspace{3cm}\caption{\label{fig:4} The scale factor versus the
temperature in the Snyder space. The scale factor has a
nonzero minimum size which leads to a nonsingular
Universe.}
\end{figure}

An important point here is that at the very high temperature, the
quantum gravitational effects dominate and the modified entropy
density (\ref{entropyNC}) changes significantly the adiabatic
condition (\ref{adiabatic}). In the usual radiation dominated era,
the entropy density is proportional to the temperature as $\propto\,
T^3$. As the temperature increases (toward the singularity of the
standard model), the usual (nondeformed) entropy density increases
and consequently the scale factor should be reduced to respect the
adiabatic condition (\ref{adiabatic}). But, the entropy density
(\ref{entropyNC}) behaves very differently in the high temperature
limit. As temperature approaches the Planck temperature, the second
term on the right hand side of the relation (\ref{entropyNC})
dominates and consequently the entropy density decreases instead of
increasing (see figure \ref{fig:3}). Interestingly, the large
entropy density problem is resolved in this way. On the other hand,
the scale factor can be obtained from the adiabatic condition
(\ref{adiabatic}) as
\begin{eqnarray}\label{SFNC}
a(T)\approx\frac{\vartheta}{T}\,\Big(1\,+\frac{\chi}{2}
(T/T_{_P})^2\Big)\,,
\end{eqnarray}
where $\vartheta^3=\frac{45\,S}{2\pi^2(g_{_{-}}+\frac{7}{8}
g_{_{+}})}$ is a numerical constant. The behavior of the scale
factor as a function of temperature is shown in figure \ref{fig:4}.

The scale factor in the standard radiation dominated era can be
obtained by setting $\chi$ to be zero in relation (\ref{SFNC}) as
$a(T)=\frac{\vartheta}{T}$ which is shown in figure \ref{fig:4} by
red dashed line. This scale factor has a singularity at the early
time or equivalently at the very high temperature. However, the
scale factor (\ref{SFNC}) has a nonzero minimum size
$a=(\sqrt{2\chi}\vartheta)\,l_{_P}$ at the temperature
$T=\sqrt{\frac{2}{\chi}}T_{_P}$. From the table \ref{tab:table1}
it is clear that this minimum occurs in the region that non of
the energy conditions are satisfied. The minimum size for the
universe when at least the WEC is satisfied happens for
temperature $T=\sqrt{\frac{3}{5\chi}}T_{_P}$ and is given by
\begin{eqnarray}\label{SFmin}
a_{\rm min}=\Big(\frac{13}{2}\sqrt{\frac{\chi}{15}}
\vartheta\Big)\,l_{_P}\,.
\end{eqnarray}
So, the Big Bang singularity is removed in this setup.

\section{Conclusions}
The Universe comes from the singularity when one uses the classical
general relativity equations to describe the cosmic evolution. It is
natural to expect that the Big Bang singularity will be removed when
one uses the yet unknown full quantum gravitational equations. While
there is no full quantum gravity theory today, some candidates such
as the string theory and loop quantum gravity revealed some aspects
of the ultimate quantum gravity theory. The minimal measurable
length and minimal momentum, which induce respectively the UV and IR
cutoffs in the corresponding theory, are the common addresses of all
promising candidates of quantum gravity proposal. In this paper we
firstly constructed a general deformed phase space with UV and IR
cutoffs by means of the deformed commutation relations. We have
shown that the Liouville theorem is satisfied in this framework
which ensures that the number of microstates remains unchanged under
the time evolution in the deformed phase space. Then we formulated a
general statistical physics which contains the UV and IR cutoffs. We
have studied the effects of the noncommutativity on the energy
density and pressure of the statistical system in the Snyder space
and we have treated the energy conditions in this framework since
now $\rho$ and $\mathrm{P}$ are not always positive definite. We
have shown that the equation of state parameter is temperature
dependent in the UV sector of the theory which opens a possible
window for the natural realization of the inflation in this setup.
The energy density of the Universe becomes finite when the DEC and
WEC are satisfied and the condition ${\mathrm{P}}=-\rho$ is possible
in the SEC, however this gives no accelerated expansion since
gravity is always attractive when SEC is satisfied. Furthermore, the
adiabatic condition $s\,a^3=\mbox{cte}$ changes in this setup since
the entropy density gets modified when one considers quantum gravity
effects. While the usual entropy density always increases as the
temperature increases, the deformed entropy density behaves very
differently in the Snyder space. The modified entropy density
decreases with temperature in the high energy regime and the
adiabatic condition implies that the scale factor gets a minimum
size as $a\propto\,l_{_P}$, where $l_{_P}$ is the Planck length.
Consequently, the Big Bang singularity will be removed in this
framework. As the final remark, we note that Einstein's equations
should be modified in the high energy regime where the
noncommutative effects are important. Some attempts have been made
in this direction (see for instance \cite{NCGCS,SZABO}), but no
complete noncommutative general relativity have been formulated yet.

\appendix

\renewcommand{\theequation}{A-\arabic{equation}}
\setcounter{equation}{0}
\section{Jacobian in Noncommutative Spaces}
In this appendix, we consider three different approaches to the
issue of the noncommutative geometry; Snyder noncommutative
spaces, coherent state approach, and the Moyal product law.

\subsection{Snyder Spaces}
We calculate the Jacobian (\ref{Jacobian}) for the Snyder algebra
(\ref{snyderalgebra}) in this section. For the small deviation from
the canonical Poisson algebra (\ref{limit}), equivalent to
$\beta,\alpha\rightarrow\,0$ in relation (\ref{snyderalgebra}), one
can use the approximate relation (\ref{Jacobianapprox}) (see
Appendix of the Ref. \cite{Fityo}). In this case, the corresponding
Jacobian becomes
\begin{eqnarray}\label{JacobianSnyder1}
J(q,p)=\underbrace{(1+|\alpha{\bf q}+\beta{\bf p}|^2)\times..
\times(1+|\alpha{\bf q}+\beta{\bf p}|^2)}_{D\,\mbox{times}}\nonumber\\
=\Big(1+|\alpha{\bf q}+\beta{\bf p}|^2\Big)^D.\hspace{0.5cm}
\end{eqnarray}
where ${\bf q}$ and {\bf p} are the $D$-vector associated to
the $q_i$ and $p_i$ respectively. Expanding above relation
in the limit of $\alpha,\beta\rightarrow\,0$, up to quadratic
order of deformation parameters, the Jacobian becomes
\begin{eqnarray}\label{JacobianSnyder2}
J(q,p)=1+\,D|\alpha{\bf q}+\beta{\bf p}|^2.
\end{eqnarray}
Having Jacobian in hand, one can study the thermodynamics of
the system through the relations (\ref{N}) and (\ref{P}).

\subsection{The Coherent States Approach}
The coherent state approach to the noncommutative geometry was
extended in Ref. \cite{NCGCS} by means of the kernels in Feynman
path integral approach in quantum field theory (see also
\cite{NCCS-Infl2}) . The noncommutativity affects the Feynman
propagator in momentum space as
\begin{equation}\label{NCGFPK}
G_{\theta}(\,p^2;\,m^2)=\frac{1}{(2\pi)^D}\,
\frac{e^{-\theta\,p^2}}{p^2+m^2},
\end{equation}
where $\theta$ is the noncommutativity parameter and
$p^2=p_1^2+p_2^2+...+p_{_D}^2$. The exponential term in the
above relation induces a UV cutoff in the high energy regime.
A more general case was extended in Ref. \cite{Nicolini} that
includes an IR cutoff as well as UV cutoff. They considered
a toy model in which the IR and UV terms appears as an
exponential term $e^{-\sigma{q}^2-\theta{p}^2}$, where
$\sigma$ is another deformation parameter that induces an IR
cutoff and $q^2=q_1^2+q_2^2+...+q_{_D}^2$. This exponential
term is effectively equivalent to the inverse of the
Jacobian in our study as
\begin{equation}\label{JacobianCS}
J^{-1}=e^{-\sigma{q}^2-\theta{p}^2}.
\end{equation}
Expanding the above relation in the limit of
$\sigma,\theta\rightarrow\,0$, up to the first order of
the deformation parameters gives
\begin{eqnarray}\label{JacobianCS2}
J(p)=1+\sigma{q^2}+\theta{p^2}+{\mathcal{O}}(
\sigma^2,\theta^2)\,,
\end{eqnarray}
which is clearly equivalent to the Snyder case
(\ref{JacobianSnyder2}) by identifying $\sigma=D\,\alpha$
and $\theta=D\,\beta^2$. However, the Jacobian
(\ref{JacobianSnyder2}) include an extra mixing term
which doesn't affect thermodynamics of the early
Universe through the relation (\ref{SnyderP2}).

\subsection{Moyal Product Law}
The last approach to the noncommutative phase space is
described by the star product, known as the Moyal
product law, between two arbitrary functions of position
and momentum as \cite{NCGCL}
\begin{eqnarray}\label{moyalproduct}
(f{\ast_{\alpha}}g)(x)=\exp\bigg(\frac{1}{2}{\alpha^{ab}}{
\partial^{(1)}_{a}}{\partial^{(2)}_{b}}\bigg)
f(x_{1})g(x_{2}){\Big{|}}_{x_{1}=x_{2}=x},\hspace{0.7cm}
\end{eqnarray}
such that
\begin{eqnarray}\label{moyalalpha}
\alpha_{ab}=\left(
              \begin{array}{cc}
                \theta_{ij} & \delta_{ij}+\sigma_{ij}\\
                -\delta_{ij}-\sigma_{ij} & \beta_{ij}\\
              \end{array}
            \right)
\end{eqnarray}
where the $D\times{D}$ matrices $\theta$ and $\beta$ are assumed to
be antisymmetric with $2D$ being the dimension of the classical
phase space, representing the noncommutativity in coordinates and
momenta, respectively. In contrast to the Poisson brackets, the
Moyal brackets can be written as
\begin{eqnarray}\label{moyalbracket}
\{f,\,g\}_{\alpha}=f\ast_{\alpha}g-g\ast_{\alpha}f\,.
\end{eqnarray}
A simple calculations shows that
\begin{eqnarray}\label{moyalalgebra}
\{q_i,q_j\}_{\alpha}=\theta_{ij},\hspace{0.2cm}\{q_i,p_j\}_{
\alpha}=\delta_{ij}+\sigma_{ij},\hspace{0.2cm}\{p_i,p_j\}_{
\alpha}=\beta_{ij}.\hspace{1cm}
\end{eqnarray}
Now, consider the transformation (\ref{transformation}) in the
classical phase space as
\begin{eqnarray}\label{poissonvariables}
(Q_i,P_i)\rightarrow{\Big(q_i=Q_i-\frac{1}{2}\theta_{ij}P_j,
\,\,\,p_i=P_i+\frac{1}{2}\beta_{ij}Q_j\Big)}.\hspace{1cm}
\end{eqnarray}
It is easy to show that the new phase space variables $q_i$ and
$p_i$ satisfy
\begin{eqnarray}\label{deformedalgebra}
\{q_i,q_j\}=\theta_{ij},\hspace{0.2cm}\{q_i,p_j\}=\delta_{ij}
+\sigma_{ij},\hspace{0.2cm}\{p_i,p_j\}=\beta_{ij},\hspace{1cm}
\end{eqnarray}
with $\sigma_{ij}=-\frac{1}{8}(\theta^{k}_{i}\beta_{kj}+
\beta^{k}_{i}\theta_{kj})$. The commutation relations
(\ref{deformedalgebra}) are the same as (\ref{moyalalgebra}). So,
transformation (\ref{transformation}) transforms commutative phase
space to the Moyal noncommutative ones. It is more convenient to
work with Poisson brackets (\ref{deformedalgebra}) than
$\alpha$-star Moyal brackets (\ref{moyalalgebra}). It is important
to note that the relations (\ref{moyalalgebra}) are defined in the
spirit of the Moyal product given above. However, in the relations
defined by (\ref{deformedalgebra}), the new phase space variables
$q_i$ and $p_i$ are functions of $Q_i$ and $P_i$ which obey the
usual Poisson bracket relations (\ref{Poisson}). So the relations
(\ref{moyalalgebra}) and (\ref{deformedalgebra}) should be
considered as distinct.

Now, we need only the Jacobian of this transformation for our
purpose. For small deviation from the Poisson algebra
($\theta_{ij}=\beta_{ij}=\sigma_{ij}\approx\,0$), the relation
(\ref{Jacobianapprox}) is a good approximation which gives the
result
\begin{eqnarray}\label{Jacobian3}
J=1\,+\,\sigma_{_{11}}+\,...\,+
\,\sigma_{_{DD}}\,=\,1\,+\,\mbox{Tr}(\sigma).
\end{eqnarray}
Note that the Jacobian becomes constant since the deformation
parameters $\theta_{ij}$, $\beta_{ij}$ and $\sigma_{ij}$ are
constant. Consequently, the Moyal noncommutativity gives no
significant modification to the thermodynamical quantities through
relations (\ref{N}) and (\ref{P}) since it multiplies the equations
just by a constant numerical factor.\\

{\bf Acknowledgement}\\
We would like to thank two anonymous referees for very insightful
comments.


\begin{thebibliography}{}
\bibitem{units} We work in natural units $\hbar=
c=k_{_B}=1$, where $\hbar$, $c$ and $k_{_B}$ are
the Planck constant, speed of light in vacuum
and Boltzman constant, respectively.
\bibitem{NCAPP} Also, there is another approach
to the issue, the so-called coherent states
approach, which directly deformes the density of
states on the phase space. However, this approach
can be effectively realized from the deformed
commutation relation (\ref{commutation}) in the
light of the Jacobian (\ref{Jacobian}) and the
corresponding deformed density of states
(\ref{QGWF}).
\end{thebibliography}
\end{document}